\documentclass[aps,prl,floats,twocolumn,showpacs,superscriptaddress,preprintnumbers,
eqsecnum,amsmath,amssymb,nofootinbib]{revtex4}

\usepackage{graphicx}
\usepackage{dcolumn}
\usepackage{bm}
\usepackage{epsfig}

\newcommand{\be}{\begin{eqnarray}}
\newcommand{\ee}{\end{eqnarray}}

\begin{document}

\preprint{
\hbox{RM3-TH/11-09}
}

\title{Matrix elements of the electromagnetic operator between kaon and pion states}

\author{I.~Baum}
\affiliation{Dip.~di Fisica, Universit{\`a} di Roma ``La Sapienza'', P.le A. Moro 5, I-00185 Roma, Italy}
\author{V.~Lubicz}
\affiliation{Dip.~di Fisica, Universit{\`a} Roma Tre, Via della Vasca Navale 84, I-00146 Roma, Italy}
\affiliation{INFN, Sezione di Roma Tre, Via della Vasca Navale 84, I-00146 Roma, Italy}
\author{G.~Martinelli}
\affiliation{SISSA, Via Bonomea 265, I-34136 Trieste, Italy}
\author{L.~Orifici}
\affiliation{Dip.~di Fisica, Universit{\`a} Roma Tre, Via della Vasca Navale 84, I-00146 Roma, Italy}
\author{S.~Simula}
\affiliation{INFN, Sezione di Roma Tre, Via della Vasca Navale 84, I-00146 Roma, Italy}

\collaboration{for the European Twisted Mass Collaboration} \noaffiliation

\begin{abstract}

We compute the matrix elements of the electromagnetic (EM) operator $\bar{s} F_{\mu \nu} \sigma^{\mu \nu} d$ 
between kaon and pion states, using lattice QCD with maximally twisted-mass fermions and two flavors of 
dynamical quarks ($N_f = 2$). 
The operator is renormalized non-perturbatively in the RI'/MOM scheme and our simulations cover pion masses 
as light as $270$ MeV and three values of the lattice spacing from $\simeq 0.07$ up to $\simeq 0.1$ fm.
At the physical point our result for the corresponding tensor form factor at zero-momentum transfer is $f_T^{K\pi}(0) 
= 0.417(14_{\rm stat})(5_{\rm syst})$,  where the systematic error does not include the effect of quenching the 
strange and charm quarks.
Our result differs significantly from the old quenched result $f_T^{K\pi}(0) = 0.78(6)$ obtained by the $\rm SPQ_{\rm cd}R$ 
Collaboration with pion masses above 500 MeV. 
We investigate the source of this difference and conclude that it is mainly related to the chiral extrapolation.
We also study the tensor charge of the pion and obtain the value $f_T^{\pi\pi}(0) = 0.195(8_{\rm stat})(6_{\rm syst})$  
in good agreement with, but more accurate than the result $f_T^{\pi\pi}(0) = 0.216(34)$ obtained by the QCDSF 
Collaboration using higher pion masses.

\end{abstract}


\maketitle

\section{Introduction}
Accurate measurements of hadron weak decays can constrain the parameters of the Standard Model (SM) and 
can place bounds on New Physics (NP) models. 
In particular, the rare decays of the kaon are ideally suited to search for new, possibly large CP-violating effects 
in the light-quark sector (see Ref.~\cite{FlaviaNet}).

In this work we present a lattice study of the matrix elements of the the electromagnetic (EM) operator between 
kaon and pion states, which may be relevant in the CP-violating part of the $K \to \pi \ell^+ \ell^-$ semileptonic 
decays.
The study has been performed using the gauge configurations generated \cite{ETMC_scaling} by the European 
Twisted Mass Collaboration (ETMC) with $N_f = 2$ maximally twisted-mass fermions \cite{TM,TM_improvement} 
and preliminary results have been presented already in Ref.~\cite{ETMC_tensor_POS}.

The EM operator involved in the weak $s \to d$ transition is given by $ \bar{s} ~ F_{\mu \nu} \sigma^{\mu \nu} d$, 
where $F_{\mu \nu}$ is the EM field tensor. 
Therefore its matrix elements between kaon and pion states involve the ones of the weak tensor current, which 
can be written in terms of a single form factor, $f_T^{K\pi}(q^2)$, as
 \be
      \langle \pi^0 | \bar{s} \sigma^{\mu \nu} d | K^0 \rangle = (p_{\pi}^\mu p_{K}^\nu - p_{\pi}^\nu p_{K}^\mu) ~
       \frac{ \sqrt{2} f_T^{K\pi}(q^2)}{M_K + M_\pi} ~ ,
      \label{eq:EM_KPi}
 \ee
where $q = (p_K - p_\pi)$ is the 4-momentum transfer and the factor $(M_K + M_\pi)^{-1}$ is conventionally inserted 
in order to make the tensor form factor dimensionless.

Our simulations cover pion masses as light as $270$ MeV and three values of the lattice spacing from $\simeq 0.07$ 
up to $\simeq 0.1$ fm.
At the physical point  our result for the $K \to \pi$ tensor form factor at zero-momentum transfer is 
 \be
     f_T^{K\pi}(0) = 0.417 ~ (14_{\rm stat}) ~ (5_{\rm syst}) = 0.417 ~ (15) ~ .
     \label{eq:fTKPi}
 \ee
where the systematic error does not include any estimate of the effect of quenching the strange and charm quarks.
Our finding differs significantly from the old quenched result $f_T^{K\pi}(0) = 0.78(6)$ obtained in Ref.~\cite{SPQCDR} 
by the $\rm SPQ_{\rm cd}R$ Collaboration with pion masses above $\sim$~500 MeV. 
The reason is mainly due to the non-analytic behavior of the tensor form factor $f_T^{K\pi}(0)$ in terms of the quark 
masses introduced by the factor $(M_K + M_\pi)^{-1}$ in the parameterization (\ref{eq:EM_KPi}). 
Such a behavior was not taken into account in Ref.~\cite{SPQCDR} (see later on).

In the case of the degenerate $\pi \to \pi$ transition, using the predictions of  the Chiral Perturbation Theory (ChPT) 
carried out in Ref.~\cite{SU2_ChPT}, we obtain for the tensor form factor $f_T^{\pi\pi}(0)$, known as the tensor charge 
of the pion, the following value 
  \be
     f_T^{\pi\pi}(0) = 0.195 ~ (8_{\rm stat})   ~ (6_{\rm syst}) = 0.195 ~ (10) ~ ,
     \label{eq:fTPiPi}
 \ee
which improves the result $f_T^{\pi\pi}(0) = 0.216(34)$ obtained by the QCDSF Collaboration \cite{QCDSF} with 
simulations at higher pion masses.

\section{$K \to \pi$ results}

We have performed the calculations of the relevant 2-point and 3-point correlation functions using the ETMC gauge 
configurations with $N_f = 2$ dynamical twisted-mass quarks generated \cite{ETMC_scaling} at three values of the 
lattice coupling $\beta$, namely the ensembles $A_2 - A_4$ at $\beta = 3.8$ ($a = 0.098(4)$ fm), $B_1 - B_7$ at 
$\beta = 3.9$ ($a = 0.085(3)$ fm), and $C_1 - C_3$ at $\beta = 4.05$ ($a = 0.067(2)$ fm). 
The pion mass $M_\pi$ ranges between $\simeq 270$ MeV and $\simeq 600$ MeV and the size $L$ of our lattices 
guarantees that  $M_\pi L$ is larger than $\sim 3.3$. 
For each value of the pion mass and of the lattice spacing we have used several values of the (bare) strange quark 
mass $m_s$ to allow for a smooth, local interpolation of our results to the physical value of $m_s$ (see 
Ref.~\cite{ETMC_masses}).
The calculation of the 2- and 3-point correlation functions has been carried out using all-to-all quark propagators evaluated 
with the {\it one-end-trick} stochastic procedure and adopting non-periodic boundary conditions which make arbitrarily small 
momenta accessible. 
All the necessary formulae can be easily inferred from Ref.~\cite{ETMC_pion}, where the degenerate case of the vector pion 
form factor is illustrated in details. 
For each pion mass the statistical errors are evaluated with the jackknife procedure.

The tensor current was renormalized non-perturbatively in the RI'/MOM scheme in Ref.~\cite{EMO_renorm}, 
including ${\cal{O}}(a^2 g^2)$ corrections evaluated at one-loop in lattice perturbation theory \cite{EMO_perturb}.
The numerical values used in our analyses for the tensor renormalization constant are $Z_T(\rm{\overline{MS}, 2 ~ GeV}) = 
0.733(9)$, $0.743(5)$, $0.777(6)$ for $\beta = 3.8, ~ 3.9, ~ 4.05$, respectively.

\begin{figure}[!htb]
\centerline{\includegraphics[width=8.5cm]{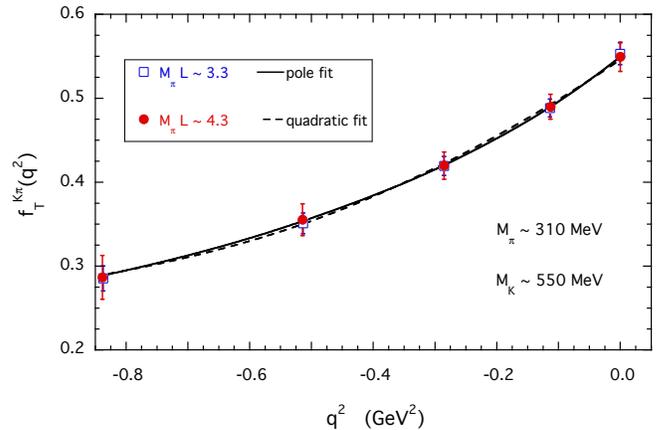}}
\caption{\label{fig:fT_q2} \it The tensor form factor $f_T^{K\pi}(q^2)$ obtained at $M_\pi \simeq 310$ MeV and 
$M_K \simeq 550$ MeV versus $q^2$ in physical units.
The dots and the squares (shifted for better clarity) correspond to the gauge ensembles $B_1$ and $B_7$, 
respectively, which differs only for the lattice size.
The solid and dashed lines are the results of the fits based on Eqs.~(\protect\ref{eq:pole}) and 
(\protect\ref{eq:quadratic}), respectively.}
\end{figure}

At each pion and kaon masses we determine the tensor form factor $f_T^{K\pi}(q^2)$ for several values of 
$q^2 < q_{max}^2 = (M_K - M_\pi)^2$ in order to interpolate at $q^2 = 0$~\footnote{We remind that the 
tensor form factor $f_T^{K \pi}(q^2)$ computed in this work is ${\cal{O}}(a)$-improved thanks to the use 
of maximally twisted Wilson quarks \cite{TM_improvement}.}. 
Note that, because of the vanishing of the Lorentz structure in Eq.~(\ref{eq:EM_KPi}), it is not possible to 
determine $f_T^{K\pi}(q^2)$ at $q^2 = q_{max}^2$.
We take advantage of the non-periodic boundary conditions to reach values of $q^2$ quite close to $q^2 = 0$.
The momentum dependence of $f_T^{K\pi}(q^2)$ can be nicely fitted either by a pole behavior 
   \be
    f_T^{K\pi}(q^2) = f_T^{K\pi}(0) / (1 - s_T^{K\pi} ~ q^2)
    \label{eq:pole}
   \ee
or by a quadratic fit in $q^2$
  \be
    f_T^{K\pi}(q^2) = f_T^{K\pi}(0) \cdot (1 + s_T^{K\pi} ~ q^2 + c_T^{K\pi} ~ q^4) ~ .
    \label{eq:quadratic}
   \ee
The good quality of both fits is illustrated in Fig.~\ref{fig:fT_q2}, where the results obtained at two different 
lattice volumes are also compared.
It can clearly be seen that: ~ i) finite size effects are well below the statistical precision of our lattice points;
~ ii) the results for $f_T^{K\pi}(0)$ and (to a less extent) for the slope $s_T^{K\pi}$, obtained using the pole 
dominance (\ref{eq:pole}), differ only slightly from those corresponding to the quadratic fit (\ref{eq:quadratic}) 
in $q^2$.

\begin{figure}[!htb]
\centerline{\includegraphics[width=8.5cm]{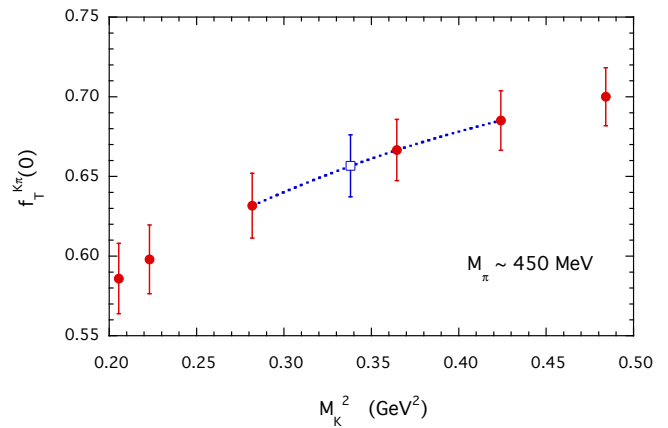}}
\caption{\label{fig:fT0} \it Results for $f_T^{K\pi}(0)$ versus $M_K^2$ at $M_\pi \simeq 450$ MeV, 
obtained assuming the pole dominance (\protect\ref{eq:pole}) for describing the $q^2$-dependence 
of our data.
The square corresponds to the value of $f_T^{K\pi}(0)$ obtained by local interpolation based 
on quadratic splines (dotted line) at the reference kaon mass $M_K^{ref} \simeq 580$ 
MeV from Eq.~(\protect\ref{eq:MKref}).}
\end{figure}

The values obtained for $f_T^{K\pi}(0)$ and $s_T^{K\pi}$ depend on both pion and kaon masses.
The dependence on the latter is shown in Fig.~\ref{fig:fT0} for $f_T^{K\pi}(0)$ at $M_\pi \simeq 450$ MeV 
and it appears to be quite smooth. 
Thus an interpolation at the physical strange quark mass can be easily performed using quadratic splines.
This is obtained by fixing the combination ($2 M_K^2 - M_\pi^2$) at its physical value, obtaining at each pion 
mass a {\it reference} kaon mass, $M_K^{ref}$, given by
 \be 
     2 [M_K^{ref}]^2 - M_\pi^2 = 2 [M_K^{phys}]^2 - [M_\pi^{phys}]^2 
     \label{eq:MKref}
 \ee
with $M_\pi^{phys} = 135.0$ MeV and $M_K^{phys} = 494.4$ MeV.

\begin{figure}[!htb]
\centerline{\includegraphics[width=8.5cm]{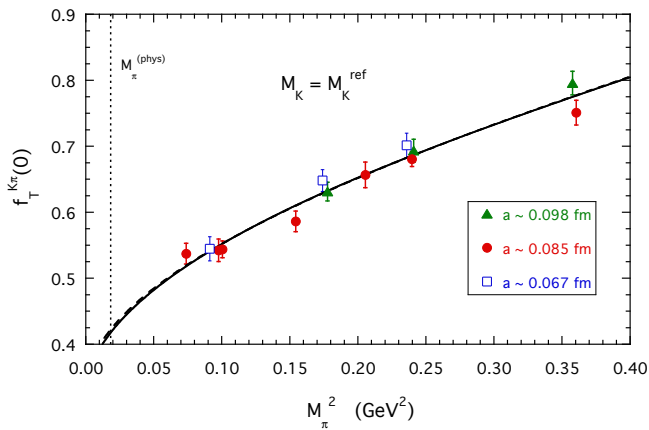}}
\caption{\label{fig:fT(0)} \it Results for $f_T^{K\pi}(0)$ versus $M_\pi^2$ at $M_K = M_K^{ref}$ in 
physical units, assuming the pole dominance (\protect\ref{eq:pole}) for describing the $q^2$-dependence 
of our data.
The dots, squares and triangles are our results for the three lattice spacings of the ETMC 
simulations, specified in the inset.
The solid and dashed lines correspond to the fit based on Eq.~(\protect\ref{eq:chiral_fit}) with 
$B = 0$ and $D = 0$, respectively.}
\end{figure}

\begin{figure}[!htb]
\centerline{\includegraphics[width=8.5cm]{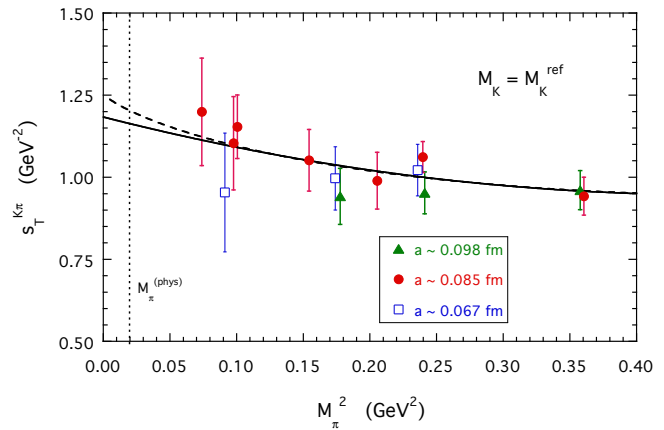}}
\caption{\label{fig:sT} \it The same as in Fig.~\protect\ref{fig:fT(0)}, but for the slope $s_T^{K\pi}$ of 
the tensor form factor at $q^2 = 0$, obtained assuming the pole dominance (\protect\ref{eq:pole}) 
for describing the $q^2$-dependence of our data.}
\end{figure}

The results for $f_T^{K\pi}(0)$ and $s_T^{K\pi}$, interpolated at the reference kaon mass $M_K^{ref}$, are 
collected in Table \ref{tab:fT_KPi} and shown in Figs.~\ref{fig:fT(0)} and \ref{fig:sT}, respectively, for the three 
lattice spacings of our simulations.
It can clearly be seen that discretization effects are sub-dominant.

\begin{table}[!htb]

\begin{center}
\begin{tabular}{||c|c|c||c|c||}
\hline
 $\mbox{Ensemble}$ & $M_\pi$  & $M_K^{ref}$ & $f_T^{K\pi}(0)$   & $f_T^{K\pi}(0)$ \\
 $(\beta,L/a)$           & (MeV)  & (MeV)  & $\mbox{(pole)}$ & $\mbox{(quadratic)}$ \\ \hline \hline
 $A_2 (3.8,24)$       & $422$ &  $569$ &  $0.631~(14)$ & $0.622~(14)$ \\ \hline 
 $A_3 (3.8,24)$       & $491$ &  $596$ &  $0.693~(17)$ & $0.674~(17)$ \\ \hline
 $A_4 (3.8,24)$       & $598$ &  $644$ &  $0.796~(18)$ & $0.783~(18)$ \\ \hline \hline
 $B_1 (3.9,24)$       & $319$ &  $534$ &  $0.544~(13)$ & $0.539~(12)$ \\ \hline
 $B_2 (3.9,24)$       & $393$ &  $559$ &  $0.586~(15)$ & $0.578~(16)$ \\ \hline
 $B_3 (3.9,24)$       & $453$ &  $581$ &  $0.657~(19)$ & $0.650~(19)$ \\ \hline
 $B_4 (3.9,24)$       & $490$ &  $596$ &  $0.681~(11)$ & $0.671~(11)$ \\ \hline
 $B_5 (3.9,24)$       & $600$ &  $645$ &  $0.751~(19)$ & $0.736~(17)$ \\ \hline \hline
 $B_6 (3.9,32)$       & $272$ &  $522$ &  $0.537~(16)$ & $0.529~(16)$ \\ \hline 
 $B_7 (3.9,32)$       & $311$ &  $533$ &  $0.542~(17)$ & $0.538~(16)$ \\ \hline \hline
 $C_1 (4.05,32)$    & $302$ &  $530$ &  $0.545~(18)$ & $0.539~(19)$ \\ \hline 
 $C_2 (4.05,32)$    & $417$ &  $568$ &  $0.648~(17)$ & $0.649~(16)$ \\ \hline
 $C_3 (4.05,32)$    & $486$ &  $594$ &  $0.702~(18)$ & $0.687~(17)$ \\ \hline \hline

\end{tabular}

\caption{\label{tab:fT_KPi} \it Results for $f_T^{K\pi}(0)$, obtained at the three lattice spacings of our simulations 
\protect\cite{ETMC_scaling}, using the pole (\protect\ref{eq:pole}) or quadratic (\protect\ref{eq:quadratic}) fits and 
interpolated at the reference kaon mass (\protect\ref{eq:MKref}) for each simulated pion mass.}

\end{center}

\end{table}

In Ref.~\cite{SPQCDR} the first lattice calculation of the EM operator matrix element between kaon 
and pion states was carried out in the quenched approximation and for pion masses above 
$\sim$~500 MeV. 
There the chiral extrapolation was performed adopting a simple linear fit of $f_T^{K\pi}(0)$ in terms 
of the squared kaon and pion masses, obtaining at the physical point the result $f_T^{K\pi}(0) = 
0.78(6)$.

In the degenerate case $M_K = M_\pi$ the chiral expansion of the tensor current  has been 
studied in Ref.~\cite{SU2_ChPT}. 
The main finding is that the form factor $f_T^{\pi\pi}(0)$ vanishes like $M_\pi$ for $M_\pi \to 0$, 
so that the ratio $f_T^{\pi\pi}(0) / M_\pi$ tends to a non-vanishing value in the chiral limit.
The same argument is expected to hold as well in the case of the $K \to \pi$ transition: the 
form factor $f_T^{K\pi}(0)$ must vanish in the SU(3) chiral limit in such a way that the ratio 
$f_T^{K\pi}(0) / (M_K + M_\pi)$ remains finite as $(M_K, ~ M_\pi) \to 0$.

Therefore we perform the chiral extrapolation of our lattice data using the Ansatz
 \be
     f_T^{K\pi}(0) & = & (M_K^{ref} + M_\pi) ~ A ~ \left[ 1 + B M_\pi^2 ~ \mbox{log}(M_\pi^2) +
                             \right. \nonumber \\
                             && \left. C M_\pi^2 + D M_\pi^4 \right] ~ ,
     \label{eq:chiral_fit}
 \ee
where $A$, $B$, $C$ and $D$ are low-energy constants, depending on the strange quark mass, 
which are kept as free parameters in our fits.
The results of the fit (\ref{eq:chiral_fit}) assuming either $B = 0$ (no chiral log) or $D = 0$ are 
shown in Fig.~\ref{fig:fT(0)} by the solid and dashed lines, respectively. 
It can be seen that the effects of the chiral log cannot be appreciated with our data.

At the physical point, after averaging the results obtained with and without the chiral log in the fitting 
function (\ref{eq:chiral_fit}) and assuming for the momentum dependence either the pole (\ref{eq:pole})
or the quadratic (\ref{eq:quadratic}) functional forms, we get 
 \be
     f_T^{K\pi}(0) = 0.417 ~ (14_{\rm stat})  \qquad \qquad \mbox{(ETMC)} ~ ,
     \label{eq:KPi_ETMC}
 \ee
where the error is statistical only.

Had we neglected the factor ($M_K^{ref} + M_\pi$) in Eq.~(\ref{eq:chiral_fit}) the result at the physical 
point would change only marginally: $f_T^{K\pi}(0) = 0.446(19_{\rm stat})$.
On the contrary, in the case of the quenched data of Ref.~\cite{SPQCDR}, which were determined 
at pion masses above $\sim$~500 MeV, the inclusion of the factor ($M_K + M_\pi$) in the chiral 
extrapolation changes significantly the result at the physical point by many standard deviations, 
namely from $f_T^{K\pi}(0) = 0.78(6)$ to
 \be
      f_T^{K\pi}(0) = 0.48 ~ (4)  \qquad \qquad {\rm (SPQ_{\rm cd}R)} ~ .
      \label{eq:KPi_SPQCDR}
 \ee
These findings indicate that the effect of the quenched approximation does not exceed $15 \%$, 
provided the correct mass factor is included in the chiral extrapolation. 

In the case of the slope $s_T^{K\pi}$ no mass factor should be considered in the chiral 
extrapolation of the lattice data shown in Fig.~\ref{fig:sT}.
After averaging the results obtained assuming for the momentum dependence either the pole 
(\ref{eq:pole}) or the quadratic (\ref{eq:quadratic}) functional forms, we get  at the physical point 
the value
\be
    s_T^{K\pi} = 1.10 ~ (8_{\rm stat}) ~ \mbox{GeV}^{-2} \qquad \qquad \mbox{(ETMC)} ~ ,
    \label{eq:sT_ETMC}
\ee
which is consistent with the quenched result $s_T^{K\pi} = 1.11(5)$ GeV$^{-2}$ from 
Ref.~\cite{SPQCDR}.

We now present our estimates of the systematic effects.

{\it Momentum dependence.} We have fitted the $q^2$-dependence of our data using either the pole 
(\ref{eq:pole}) or the quadratic (\ref{eq:quadratic}) fits (see Fig.~\ref{fig:fT_q2}). 
We have found a systematic error due to these different choices equal to $0.002$ for $f_T^{K\pi}(0)$ 
and $0.09$ GeV$^{-2}$ for $s_T^{K\pi}$.

{\it Finite Size.} The comparison between the simulations corresponding to the gauge ensembles 
$B_1$ and $B_7$ (see Table \ref{tab:fT_KPi} and Fig.~\ref{fig:fT_q2}), which differs only for the 
lattice size, indicates a very small volume effect of the order of $0.4 \%$ on $f_T^{K\pi}(0)$ and 
of $4 \%$ on $s_T^{K\pi}$ at $M_\pi \sim 310$ MeV. 
We quote therefore a systematic error due to finite size effects equal to $0.002$ for $f_T^{K\pi}(0)$ 
and $0.05$ GeV$^{-2}$ for $s_T^{K\pi}$.

{\it Discretization.} The results obtained using only the data at a single lattice spacing $a \sim 0.085$ 
fm, where the number of simulated pion masses ensures the most reliable chiral extrapolation, are 
$f_T^{K\pi}(0) = 0.420(10_{\rm stat})$ and $s_T^{K\pi} = 1.12(6_{\rm stat})$.
We estimate therefore a systematic error due to discretization effects equal to $0.003$ for 
$f_T^{K\pi}(0)$ and $0.02$  GeV$^{-2}$ for $s_T^{K\pi}$.

{\it Chiral extrapolation.} As shown in Figs.~\ref{fig:fT(0)}-\ref{fig:sT}, the difference in the extrapolations 
to the physical point including or excluding a chiral log in the fitting function (\ref{eq:chiral_fit}) is quite 
small, being equal to $0.002$ for $f_T^{K\pi}(0)$ and $0.02$  GeV$^{-2}$ for $s_T^{K\pi}$. 
We take these numbers as our estimates of the uncertainty due to the chiral extrapolation.

Adding all the systematic errors in quadrature, our final results are
 \be
      \label{eq:fT(0)_final}
      f_T^{K\pi}(0) & = & 0.417 ~ (14_{\rm stat}) ~ (5_{\rm syst}) \nonumber \\
                             & = & 0.417 ~ (15) ~ , \\[4mm]
      \label{eq:sT_final}
      s_T^{K\pi} & = & 1.10 ~ (8_{\rm stat}) ~ (11_{\rm syst}) ~ \mbox{GeV}^{-2} \nonumber \\
                         & = & 1.10 ~ (14) ~ \mbox{GeV}^{-2} ~ ,
 \ee
where the systematic errors do not include any estimate of the effect of quenching the strange and 
charm quarks. 

\subsection{Bound on the supersymmetric coupling $\delta_+$}

We want now to use our new determination (\ref{eq:fT(0)_final}) for the tensor form factor $f_T^{K\pi}(0)$ as 
well as the recent experimental bound \cite{KTEV} on the branching ratio ${\cal{B}}(K_L \to \pi^0 e^+ e^-)$ 
to update the bound on the supersymmetric coupling $\delta_+$, related to the splitting of the off-diagonal 
entries in the down-type squark mass matrix, already calculated in Ref.~\cite{SPQCDR} using the quenched 
estimate $f_T^{K\pi}(0) = 0.78(6)$.

The master formula, expressing the CP violating part of the contribution of the EM operator to the rare 
decay $K_L \to \pi^0 e^+ e^-$, can be written as (see Ref.~\cite{SPQCDR})
 \be
     {\cal{B}}(K_L \to \pi^0 e^+ e^-) = 5.3 \cdot 10^{-4} ~ B_T^2 ~ (\mbox{Im}~\delta_+)^2 ~ ,
     \label{eq:BR}
  \ee
where the numerical coefficient ($5.3 \cdot 10^{-4}$) is the appropriate one for the EM operator renormalized 
in the $\overline{MS}$ scheme at a renormalization scale of $2$ GeV and for a gluino and average squark 
masses of $500$ GeV.
In Eq.~(\ref{eq:BR}) the quantity $B_T$ is the main hadronic parameter related to the tensor form factor at 
zero-momentum transfer $f_T^{K\pi}(0)$ by
 \be
    B_T = \frac{2 M_K}{M_K + M_\pi} \frac{f_T^{K\pi}(0)}{f_+^{K\pi}(0)} \cdot T ~ ,
    \label{eq:BT}
 \ee
where $f_+^{K\pi}(0)$ is the vector form factor at zero-momentum transfer, appearing in the $K_{\ell 3}$ semileptonic 
decay $K^+ \to \pi^0 e^+ \nu_e$, while $T$ is a correction which takes into account the different $q^2$-dependence 
of the vector and the tensor form factors (see Ref.~\cite{SPQCDR} for its definition).
The quantity $T$ can be related to the slopes $s_+^{K\pi} $ and $s_T^{K\pi}$ of the vector and tensor form factors at 
$q^2 = 0$ by the following relation
 \be 
    T = 1 - 2 M_\pi^2 \left( s_+^{K\pi} - s_T^{K\pi} \right) R ~ ,
    \label{eq:T}
 \ee
where $R$ is a simple kinematical factor depending on the kaon and pion masses.
At the physical point one has $R = 1.855$.

In Ref.~\cite{Kl3_ETMC} we investigated the vector and scalar form factors entering the $K_{\ell 3}$ semileptonic decay, 
obtaining the results $f_+^{K\pi}(0) = 0.9560 (57_{\rm stat}) (62_{\rm syst})$ and $s_+^{K\pi} = 1.30 (13_{\rm stat})  
(12_{\rm syst})$ GeV$^{-2}$.
Using the latter values and the results (\ref{eq:fT(0)_final}-\ref{eq:sT_final}), the correction factor $T$ turns out to be quite close 
to unity, namely $T = 0.986 (10_{\rm stat})  (11_{\rm syst})$, and for the quantity $B_T$ we get 
 \be
    B_T = 0.676 ~ (24_{\rm stat}) ~ (12_{\rm syst}) = 0.676 ~ (27) ~ .
    \label{eq:BT_final}
 \ee
Therefore, using in Eq.~(\ref{eq:BR}) the experimental upper bound ${\cal{B}}(K_L \to \pi^0 e^+ e^-) < 2.8 \cdot 10^{-10}~(90\%~\mbox{C.L.})$ 
\cite{KTEV} and our result (\ref{eq:BT_final}), we obtain the bound
 \be
    |\mbox{Im} ~ \delta_+| < 1.1 \cdot 10^{-3} \qquad (90\%~\mbox{C.L.}) ~ .
 \ee

\section{Pion tensor charge}

Following Ref.~\cite{SU2_ChPT} the chiral expansion of the pion tensor charge $f_T^{\pi\pi}(0)$ 
has the form
 \be
     f_T^{\pi\pi}(0) & = & M_\pi ~ A' ~ \left[ 1 + \frac{M_\pi^2}{(4 \pi f)^2} ~ \mbox{log}(M_\pi^2) + 
                              \right. \nonumber \\
                              && \left. C' M_\pi^2 + D' M_\pi^4 \right] ~ ,
     \label{eq:chiral_fit_pion}
 \ee
where $f  \simeq 122$ MeV \cite{ETMC_masses} is the pion decay constant in the SU(2) chiral limit and 
the presence of the factor $M_\pi$ is expected to have an important, bending effect on the value 
extrapolated to the physical point.

\begin{figure}[!htb]
\centerline{\includegraphics[width=8.5cm]{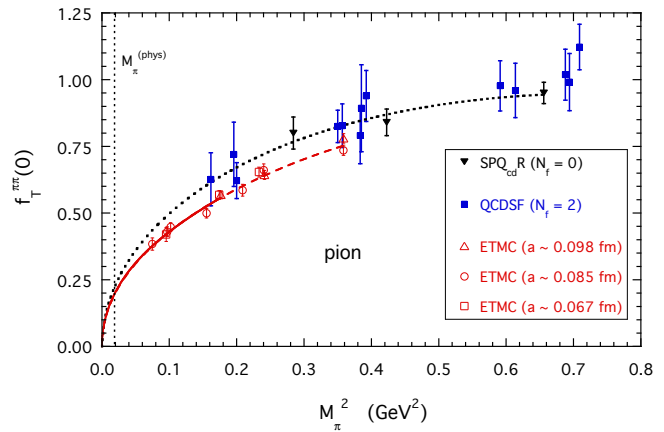}}
\caption{\label{fig:pion} \it Results for the pion tensor charge $f_T^{\pi\pi}(0)$ versus $M_\pi^2$ in physical units 
from our simulations at the three different values of the lattice spacing (open markers),  obtained assuming the 
pole dominance (\protect\ref{eq:pole}) for describing the $q^2$-dependence of our data, and from 
Refs.~\protect\cite{SPQCDR} (full triangles) and \protect\cite{QCDSF} (full squares).
The dashed line is the result of the fit (\protect\ref{eq:chiral_fit_pion}) with $D' \neq 0$ applied to all ETMC points, 
while the solid line corresponds to the use of the fitting function (\protect\ref{eq:chiral_fit_pion}) with $D' = 0$ applied 
to the ETMC data with $M_\pi \lesssim 420$ MeV.
The dotted line represents the fit described in the text and applied to the quenched data of Ref.~\protect\cite{SPQCDR}.}
\end{figure}

Our results for $f_T^{\pi\pi}(0)$, obtained at three values of the lattice spacing and for $270~\rm{MeV} \lesssim 
M_\pi \lesssim 600~\rm{MeV}$, are shown in Fig.~\ref{fig:pion} and compared with the $N_f = 2$ results of 
Ref.~\cite{QCDSF}, having $M_\pi \gtrsim 440$ MeV, and with the quenched calculations of Ref.~\cite{SPQCDR}, 
ranging from $M_\pi \sim 530$ MeV up to $M_\pi \sim 800$ MeV.
It can be seen that our results have a much better statistical precision and cover much lighter pion masses, 
where the bending effect due to the overall factor $M_\pi$ is clearly visible.

\begin{table}[!htb]

\begin{center}
\begin{tabular}{||c|c||c|c||}
\hline
 $\mbox{Ensemble}$ & $M_\pi$  & $f_T^{\pi\pi}(0)$   & $f_T^{\pi\pi}(0)$ \\
 $(\beta,L/a)$           & (MeV)  & $\mbox{(pole)}$ & $\mbox{(quadratic)}$ \\ \hline \hline
 $A_2 (3.8,24)$       & $422$ & $0.568~(15)$ & $0.555~(13)$ \\ \hline 
 $A_3 (3.8,24)$       & $491$ & $0.643~(17)$ & $0.623~(16)$ \\ \hline
 $A_4 (3.8,24)$       & $598$ & $0.780~(18)$ & $0.767~(17)$ \\ \hline \hline
 $B_1 (3.9,24)$       & $319$ & $0.448~(16)$ & $0.432~(14)$ \\ \hline
 $B_2 (3.9,24)$       & $393$ & $0.500~(18)$ & $0.487~(18)$ \\ \hline
 $B_3 (3.9,24)$       & $453$ & $0.586~(22)$ & $0.580~(20)$ \\ \hline
 $B_4 (3.9,24)$       & $490$ & $0.661~(24)$ & $0.641~(22)$ \\ \hline
 $B_5 (3.9,24)$       & $600$ & $0.736~(19)$ & $0.720~(18)$ \\ \hline \hline
 $B_6 (3.9,32)$       & $272$ & $0.384~(23)$ & $0.371~(20)$ \\ \hline 
 $B_7 (3.9,32)$       & $311$ & $0.431~(18)$ & $0.426~(15)$ \\ \hline \hline
 $C_1 (4.05,32)$    & $302$ & $0.420~(26)$ & $0.404~(23)$ \\ \hline 
 $C_2 (4.05,32)$    & $417$ & $0.569~(16)$ & $0.562~(15)$ \\ \hline
 $C_3 (4.05,32)$    & $486$ & $0.654~(19)$ & $0.640~(17)$ \\ \hline \hline

\end{tabular}

\caption{\label{tab:fT_pion} \it Results for $f_T^{\pi\pi}(0)$, obtained at the three lattice spacings of our simulations 
\protect\cite{ETMC_scaling}, using the pole or quadratic fits for describing the $q^2$-dependence of our data.}

\end{center}

\end{table}

Applying Eq.~(\ref{eq:chiral_fit_pion}) with $D' = 0$ to our lattice points  with $M_\pi \lesssim 420$ MeV (see the solid 
line in Fig.~\ref{fig:pion}), we get at the physical point $f_T^{\pi\pi}(0) = 0.195(8_{\rm stat})$.

As for the estimates of the systematic effects, we follow the same procedures described in the previous Section 
for the $K \to \pi$ transition, obtaining an effect equal to $0.004$ from fitting the momentum dependence with either a 
pole or a quadratic forms, $0.003$ from the finite size\footnote{The comparison between the simulations corresponding 
o the gauge ensembles $B_1$ and $B_7$, which differs only for the lattice size, indicates a volume effect of the order of 
$2 \div 4 \%$ on $f_T^{\pi\pi}(0)$ (see Table \protect\ref{tab:fT_pion}), which is close to the statistical error. Such an effect 
is however mostly due to the volume effect on the pion mass. Indeed, the finite size effect on the ratio $f_T^{\pi\pi}(0) / 
M_\pi$ reduces to $\simeq 1.5 \%$, which we therefore take as our estimate of the corresponding systematic error 
on our final result.}, $0.001$ from discretization and $0.003$ from the chiral extrapolation.
Adding all the systematic errors in quadrature, our final result is 
 \be
     f_T^{\pi\pi}(0) & = & 0.195 ~ (8_{\rm stat}) ~ (6_{\rm syst}) \nonumber \\
     & = & 0.195 ~ (10) \qquad \qquad \qquad \mbox{(ETMC)} 
     \label{eq:pion_ETMC}
 \ee
which improves  the QCDSF result \cite{QCDSF}  
 \be
     f_T^{\pi\pi}(0) = 0.216 ~ (34) \qquad \qquad \mbox{(QCDSF)} ~ . 
     \label{eq:pion_QCDSF}
 \ee
Finally, we also apply a simple fit of the form $f_T^{\pi\pi}(0) = M_\pi ~ A' ~ \left[ 1 + C' M_\pi^2 \right]$ 
to the three quenched data of Ref.~\cite{SPQCDR} (see the dotted line in Fig.~\ref{fig:pion}), obtaining at the physical 
point the result
 \be
     f_T^{\pi\pi}(0) = 0.221~(21_{\rm stat}) \qquad \qquad {\rm (SPQ_{\rm cd}R)} ~ , 
     \label{eq:pion_SPQCDR}
 \ee
which clearly shows that quenching effects on the pion tensor charge are sub-dominant  with respect to the present 
precision.

Before closing we want to comment briefly on the $SU(3)$ breaking effect on the tensor form factor at zero-momentum 
transfer.
Our final values obtained for $f_T^{K\pi}(0)$ and $f_T^{\pi\pi}(0)$ [i.e., Eqs.~(\ref{eq:fT(0)_final})  and (\ref{eq:pion_ETMC})] 
differ by a factor of $\approx 2$.
This huge difference is however mostly due to the factor $(M_K + M_\pi)$, conventionally inserted in Eq.~(\ref{eq:EM_KPi}).
By introducing the quantities
 \be
    F_T^{K\pi}(0) &  \equiv & f_T^{K\pi}(0) / (M_K + M_\pi) \\
    F_T^{\pi\pi}(0) & \equiv & f_T^{\pi\pi}(0) / 2 M_\pi
 \ee
one gets from  Eqs.~(\ref{eq:fT(0)_final})  and (\ref{eq:pion_ETMC}) the values
 \be
    F_T^{K\pi}(0) & = & 0.663 ~ (22_{\rm stat}) ~ (8_{\rm syst}) ~ \mbox{GeV}^{-1} \nonumber \\
                             & = & 0.663 ~ (24) ~ \mbox{GeV}^{-1}\\
    F_T^{\pi\pi}(0) & \equiv & 0.722 ~ (30_{\rm stat}) ~ (22_{\rm syst}) ~ \mbox{GeV}^{-1} \nonumber \\
                               & = & 0.722 ~ (37) ~ \mbox{GeV}^{-1}
 \ee
which yield a ratio $F_T^{K\pi}(0) / F_T^{\pi\pi}(0)$ equal to $0.92 (6)$.
This indicates that the $SU(3)$ breaking effect on the matrix element of the EM operator $\bar{s} F_{\mu \nu} \sigma^{\mu \nu} d$ 
at zero-momentum transfer is equal to $-8 (6)\%$.
This value is of the same sign and order of magnitude of the $SU(3)$ breaking effect on the corresponding matrix element of the 
vector current, namely on the vector form factor at zero-momentum transfer $f_+^{K\pi}(0)$.
Indeed, taking into account that $f_+^{\pi\pi}(0) = 1$ because of charge conservation and $f_+^{K\pi}(0) = 0.9560 (84)$ from 
Ref.~\cite{Kl3_ETMC}, the SU(3) breaking effect on the matrix element  of the vector current at zero-momentum transfer is equal to 
$-4 (1) \%$.

\section*{Acknowledgements}
We thank the ETMC members for fruitful discussions and the apeNEXT computer centre in Rome for its invaluable technical help.


\begin{thebibliography}{99}

\bibitem{FlaviaNet}
  M.~Antonelli, D.~M.~Asner, D.~A.~Bauer, T.~G.~Becher, M.~Beneke, A.~J.~Bevan, M.~Blanke, C.~Bloise {\it et al.},
  Phys.\ Rept.\  {\bf 494 } (2010) 197-414.
  [arXiv:0907.5386 [hep-ph]].

\bibitem{ETMC_scaling}
  R.~Baron {\it et al.}  [ETM Collaboration],
  JHEP {\bf 1008} (2010) 097
  [arXiv:0911.5061 [hep-lat]].

  Ph.~Boucaud {\it et al.}  [ETM Collaboration],
  Phys.\ Lett.\  B {\bf 650} (2007) 304
  [arXiv:hep-lat/0701012].

  Ph.~Boucaud {\it et al.}  [ETM Collaboration],
  Comput.\ Phys.\ Commun.\  {\bf 179} (2008) 695
  [arXiv:0803.0224 [hep-lat]].

\bibitem{TM}
  R.~Frezzotti {\it et al.} [Alpha Collaboration],
  JHEP {\bf 0108} (2001) 058
  [hep-lat/0101001].
  
\bibitem{TM_improvement}
  R.~Frezzotti, G.~C.~Rossi,
  JHEP {\bf 0408} (2004) 007
  [hep-lat/0306014].
  
\bibitem{ETMC_tensor_POS}
  I.~Baum, V.~Lubicz, G~Martinelli  and S.~Simula,
  PoS {\bf LATTICE2010 } (2010)  297.
  
\bibitem{SPQCDR}
  D.~Becirevic, V.~Lubicz, G.~Martinelli and F.~Mescia [SPQcdR
  Collaboration],
  Phys. Lett. {\bf B501}, 98 (2001)
  [arXiv:hep-ph/0010349].

\bibitem{SU2_ChPT}
  M.~Diehl, A.~Manashov, A.~Schafer,
  Phys.\ Lett.\  {\bf B622} (2005)  69-82
  [hep-ph/0505269];
  Eur.\ Phys.\ J.\  {\bf A31} (2007)  335-355
  [hep-ph/0611101].

\bibitem{QCDSF}
  D.~Brommel {\it et al.}  [QCDSF and UKQCD Collaborations],
  Phys.\ Rev.\ Lett.\  {\bf 101}, 122001 (2008)
  [arXiv:0708.2249 [hep-lat]].

\bibitem{ETMC_masses}
  B.~Blossier {\it et al.} [ETM Collaboration],
  Phys.\ Rev.\  {\bf D82} (2010) 114513
  [arXiv:1010.3659 [hep-lat]];
  PoS {\bf LATTICE2010 } (2010) 239
  [arXiv:1011.1862 [hep-lat]].

\bibitem{ETMC_pion}
  R.~Frezzotti {\it et al.} [ETM Collaboration],
  Phys.\ Rev.\  {\bf D79} (2009) 074506
  [arXiv:0812.4042 [hep-lat]].

\bibitem{EMO_renorm}
  M.~Constantinou {\it et al.} [ETM Collaboration],
  JHEP {\bf 1008}, 068 (2010)
  [arXiv:1004.1115 [hep-lat]].

\bibitem{EMO_perturb}
  M.~Constantinou, V.~Lubicz, H.~Panagopoulos and F.~Stylianou,
  JHEP {\bf 0910}, 064 (2009)
  [arXiv:0907.0381 [hep-lat]].

\bibitem{KTEV}
 A.~Alavi-Harati {\it et al.} [KTeV Collaboration],
  Phys.\ Rev.\ Lett.\  {\bf 93} (2004) 021805
  [hep-ex/0309072].

\bibitem{Kl3_ETMC}
 V.~Lubicz {\it et al.} [ETM Collaboration],
  Phys.\ Rev.\  {\bf D80} (2009) 111502
  [arXiv:0906.4728 [hep-lat]].
  
  See also V.~Lubicz {\it et al.} [ETM Collaboration],
  PoS {\bf LATTICE2010} (2010) 316
  [arXiv:1012.3573 [hep-lat]].

\end{thebibliography}
\end{document}